# Universal Quantum Gate Compilation in SU(2)$_k$ Anyon Models via Multiple-Braiding

Jiangwei Long [1], Zihui Liu [1], Yizhi Li [2] and Lijun Meng [1, †]

[1] School of Physics and Optoelectronics, Xiangtan University, Xiangtan 411105, Hunan, People's Republic of China

[2] School of Physics and Electronic Science, Hunan Institute of Science and Technology, Yueyang 414006, People's Republic of China

We investigate the capability of multiple-braiding in SU(2)$_k$ anyon models to realize universal quantum computation. The multiple elementary braiding matrices (MEBMs) are derived from the $q$-deformed representation theory of SU(2). Through a general analysis of the MEBMs, we find that multiple-braiding loses its universality only when the braiding multiplicity $m$ causes the MEBMs to collapse to a scalar (up to a global phase). Our numerical analysis of the MEBMs of SU(2)$_k$ anyon models with $k$ = 3, 5, 6, 7, and $m \in [2, 9]$, shows that the values of $m$ at which the braid group representation fails to be dense agree exactly with the theoretical prediction. For those $m$ that support universality, single-qubit gates are compiled to high precision by a genetic algorithm-enhanced Solovay–Kitaev algorithm (GA-enhanced SKA), and a genetic algorithm (GA) search with progressively increasing braid length yields an approximately the local equivalence class [CNOT]. Notably, even-order braiding operations offer a physical advantage by reducing the number of non-Abelian anyons required in braiding-based topological quantum computation (TQC). Our analytical and numerical results together provide strong evidence for identifying which braiding multiplicities $m$ support universal quantum computation in SU(2)$_k$ anyon models.

## I Introduction

Although quantum computing has demonstrated remarkable advantages over classical computing on certain specific problems [1], it faces the challenge of decoherence due to susceptibility to environmental noise. Topological quantum computation (TQC) encodes quantum information in robust ground-state degeneracies, thereby providing inherent global protection against local perturbations and emerging as a highly promising paradigm [2]. TQC relies on quasiparticle excitations known as anyons in two-dimensional systems. The idea of performing TQC with non-Abelian anyons was first proposed by Kitaev [3]. While the experimental detection of non-Abelian anyons remains challenging, encouraging progress has been made in topological superconductors [4] and fractional quantum Hall systems [5] toward realizing such exotic particles.

The SU(2)$_k$ anyon models describe a significant family of topological phases [6]. In particular, the potential for universal quantum computation offered by this series of models (specifically for $k > 2$, $k \neq 4$) has motivated substantial research interest, and their computational universality has been established both theoretically [7] and numerically [8].

In contrast to the SU(2)$_2$ (Ising) and SU(2)$_4$ (metaplectic) anyon models, where standard quantum gates can be implemented directly with only a few braiding operations [9-15], constructing standard gates in $SU(2)_k$ models (with $k > 2$ and $k \neq 4$) requires approximating a target unitary operations with increasingly long sequences of elementary braids [8]. This process corresponds to synthesizing a braidword whose overall action approximates a desired quantum gate (up to a global phase). As the length (the number of braiding operations in a braidword) increases, the number of possible sequences grows exponentially, giving rise to the topological quantum compilation problem [16]. Solutions to this problem have been extensively studied [17-21]. While these methods were initially developed for compiling single-qubit gates in the $SU(2)_3$ (Fibonacci) anyon model, they can be readily generalized to other $SU(2)_k$ models with $k > 4$. Recently, a new type anyon *neglecton* has been introduced into the $SU(2)_2$ anyon model. The resulting braiding matrices and fusion rules enable $SU(2)_2$ anyons to achieve universal quantum computation through braiding alone, without any additional operations [22,23]. Furthermore, a universal quantum gate set constructed from the braiding operations of the non-semisimple $SU(2)_2$ anyon model has been demonstrated numerically [24].

† Corresponding author. E-mail: ljmeng@xtu.edu.cn

For construction two-qubit entangling gates of the $SU(2)_3$ anyon model, initial proposals relied on the unique fusion rules ( $\tau \otimes \tau = 1 \oplus \tau$ , where $\tau$ denotes Fibonacci anyon, and 1 denotes vacuum) of the model to implement the controlled-injection approach [25]. An alternative approach approximates the local equivalence class [CNOT] gate using the elementary braiding matrices (EBMs) of the two-qubit encoding [26], this method that also extends to $SU(2)_k$ models with $k > 4$ [8]. Furthermore, by leveraging cabling concept from knot theory, specific braid sequences can be selected to achieve low-error entangling gates [27]. While the controlled-injection approach based on three-qubit [28] and even $N$-qubit constructions [29] have been proposed for $SU(2)_3$ anyons.

In braiding-based TQC, logical gates are typically realized by sequences of elementary braids involving two anyons at a time. It has been theoretically established that double-braiding within $SU(2)_k$ ( $k > 2$, $k \neq 4$) anyon models is dense in SU(2) [30]. While previous work has demonstrated that universal quantum computation can be achieved by moving a single anyon via injection braiding, this approach requires the insertion of braidwords equivalent to the identity operation to control the positions of anyons, thereby incurring additional braiding steps [31]. In contrast, double-braiding achieves the same reduction in anyon manipulation without introducing such extra operations. Such braiding has also been employed to retrieve the Chern–Simons level $k$ experimentally, by measuring the annihilation probability of an anyon pair after repeated windings [32].

In this work, double-braiding operations are extended to multiple-braiding (the consecutive application of multiple identical braid operations treated as a single effective unit). We have derived the general form of the multiple elementary braiding matrices (MEBMs) by composing the $F$ (fusion) and $R$ (rotation) moves, evaluated them explicitly using the $q$-deformed representation theory of SU(2). Based on this general form, we have determined precisely for which values of the braiding multiplicity $m$ the multiple braiding operations are dense on the Bloch sphere. We have further numerically compiled the universal gate set {$H$, $T$, CNOT} from the MEBMs of the SU(2)$_k$ anyon models, with the numerical results in complete agreement with our theoretical predictions. Finally, by providing braidwords approximating {$H$, $T$, CNOT}, we have demonstrated how even-order braiding operations reduce the number of anyons that must be physically controlled.

The paper is structured as follows: In Sec. II, we introduce the SU(2)$_k$ anyon models and present the theoretical prediction for the denseness of the MEBMs. Section III reports the numerical results of gate compilation via multiple-braiding in the SU(2)$_3$ model, together with a performance comparison between single-braiding ($m$ = 1) and multiple-braiding ($m \in [2, 9]$). Section IV concludes with a summary. Appendix A provides the explicit form of the MEBM $\left(\sigma_3^{(6)}\right)^m$ in the SU(2)$_k$ anyon models. Appendix B presents the numerical results of gate compilation via multiple-braiding in the SU(2)$_5$, SU(2)$_6$, and SU(2)$_7$ models. Appendix C illustrates how, within the even-order braiding framework, topological equivalence can be exploited to minimize the number of anyons that must be actively controlled.

## II Theoretical Framework and Computational Methods

In SU(2)$_k$ anyon models, distinct anyon types are labeled by topological spins ranging from 0 to $k$/2 in steps of 1/2. Their fusion rules are governed by the associated $k$-level theory [33]:

$$j_1 \otimes j_2 = \bigoplus_{j=|j_1-j_2|}^{\min\{j_1+j_2,\, k-j_1-j_2\}} j, \tag{1}$$

where $j_1$ and $j_2$ denote the topological spins of the two anyons to be fused, and $j$ represents the topological spin of the fusion outcome. The fusion can yield multiple possible values of $j$ for non-Abelian anyons. The symbol $\otimes$ denotes the fusion operation, while $\oplus$ indicates the multiplicity of a possible fusion channel.

One qubit can be encoded using either three or four identical anyons [34], provided their fusion exhibits a twofold degenerate channel. For SU(2)$_k$ models with k ≥ 5, a suitable encoding is realized with three anyons carrying 1/2-topological spin [8]. In the case of SU(2)$_3$ (Fibonacci anyons), the anyon types are effectively reduced from {0, 1/2, 1, 3/2} to {0, 1} (see Fig. 2 of Ref. [16]). Consequently, one qubit in the SU(2)$_3$ model is constructed from three anyons with 1-topological spin.

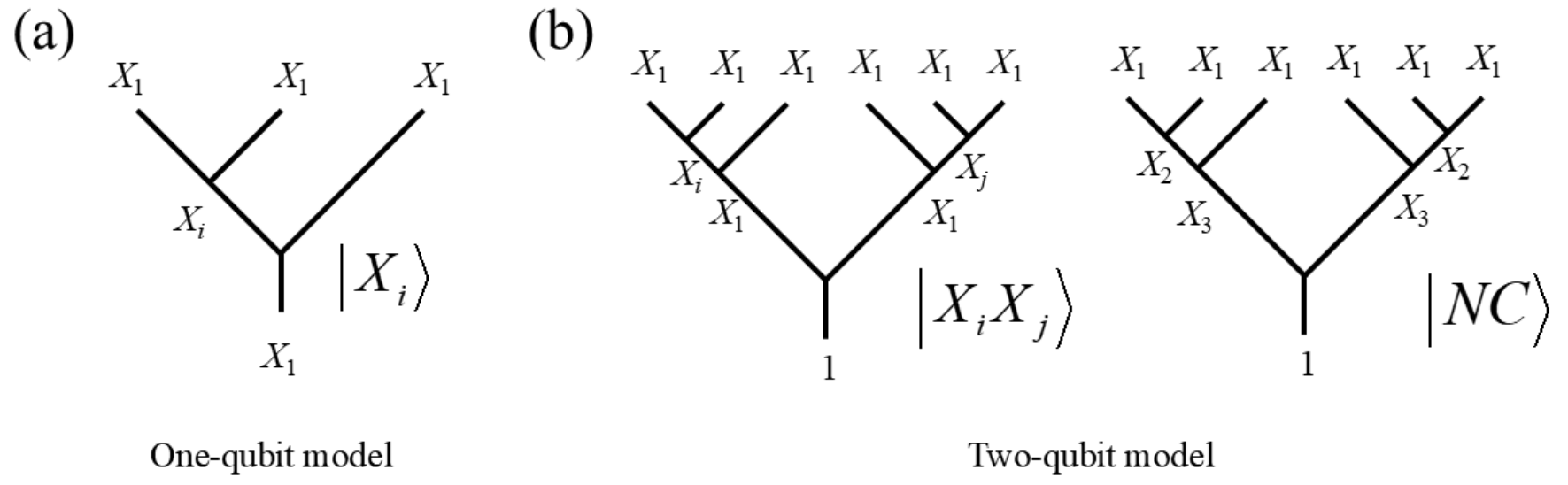

FIG. 1. Schematic of qubit encoding using anyons. (a) A single-qubit is formed by three anyons with 1/2-topological spin. (b) A two-qubit system is composed of six anyons with 1/2-

topological spin. Here, the first group of anyons (left) encodes the computational basis states, while the second group (right) resides in a non-computational state.

In the SU(2)$_k$ ($k \geq 5$) anyon models, the qubit can be encoded as illustrated in Fig. 1. For clarity, we label distinct anyon types by double their topological spin written as a subscript—e.g., $X_1$ denotes an anyon with 1/2-topological spin. As shown in Fig. 1(a), three anyons of type $X_1$ encode a single logical qubit. The fusion of the first and second $X_1$ yields an intermediate charge, which can be either the vacuum **1** or the anyon $X_2$. These two possibilities $\{|\mathbf{1}\rangle, |X_2\rangle\}$ correspond to the computational basis states $|0\rangle$ and $|1\rangle$, respectively. The intermediate charge then fuses with the third $X_1$ to give a final anyon of type $X_1$. Fig. 1(b) illustrates the two-qubit encoding using six $X_1$ anyons. The first and second anyons fuse to an intermediate charge $c_1 \in \{\mathbf{1}, X_2\}$, while the fifth and sixth anyons fuse to $c_2 \in \{\mathbf{1}, X_2\}$. The four combinations $|c_1, c_2\rangle$ $\{|\mathbf{1},\mathbf{1}\rangle, |\mathbf{1}, X_2\rangle, |X_2,\mathbf{1}\rangle, |X_2, X_2\rangle\}$ correspond to the two-qubit basis states $\{|0,0\rangle, |0,1\rangle, |1,0\rangle, |1,1\rangle\}$. These intermediate charges subsequently fuse with the third and fourth $X_1$ anyons, resulting in two final anyons of type $X_1$. Finally, the fusion of these two $X_1$ anyons yields the overall vacuum charge **1**. Throughout this process, a non-computational state $|NC\rangle$ may also appear.

For the SU(2)$_3$ (Fibonacci) anyon model, the analogous one- and two-qubit encodings are described in Ref. [16] (Fig. 4) and Ref. [35] (Fig. 3), respectively.

Using the encoding scheme shown in Fig. 1(a) with the computational basis $\{|\mathbf{1}\rangle, |X_2\rangle\}$ in the SU(2)$_k$ ($k \geq 5$) anyon models with, the single-qubit MEBM can be summarized as follows:

$$\left(\sigma_1^{(3)}\right)^m = \begin{bmatrix} \left(R_0^{11}\right)^m & 0 \\ 0 & \left(R_2^{11}\right)^m \end{bmatrix}, \tag{2}$$

$$\left(\sigma_2^{(3)}\right)^m = F\left(\sigma_1^{(3)}\right)^m F^{-1}, F = \begin{bmatrix} F_{1;00}^{111} & F_{1;20}^{111} \\ F_{1;02}^{111} & F_{1;22}^{111} \end{bmatrix}, \tag{3}$$

where $R$-symbols encode the phase acquired when two identical anyons are exchanged by 180° rotation, while the $F$-matrices describe the transformation between different fusion bases, with their entries corresponding to specific recoupling coefficients. Their explicit values can be derived from the $q$-deformed representation theory of SU(2). All necessary $F$-matrices and $R$-symbols used in our calculations are provided in the references [8]. The notation $\left(\sigma_i^{(3)}\right)^m$ (with $i$ = 1, 2) denotes $m$ successive braiding operations between the $i$-th and ($i$+1)-th anyons within the 3-anyon single-qubit encoding.

For the three-anyon single-qubit encoding, the structures of the two elementary braid generators, Eqs. (2) and (3), imply that conjugation by the $F$-matrix leaves the eigenvalues of $\left(\sigma_1^{(3)}\right)^m$ invariant. Rewriting Eq. (2) as:

$$\left(\sigma_1^{(3)}\right)^m = \left(R_2^{11}\right)^m \begin{bmatrix} (\lambda)^m & 0 \\ 0 & 1 \end{bmatrix}, \tag{4}$$

where $\lambda = R_0^{11} / R_2^{11}$, Eq. (4) reveals that the generators of the multiple-braiding depend on the the eigenvalue ratio $\lambda$ and $m$. Consequently, whether the $m$-fold multiple braiding is dense on the Bloch sphere reduces to whether $\lambda^m$ degenerates to $\pm 1$: the MEBMs collapse (up to a global phase) into a scalar ($\lambda^m = +1$) or a Pauli $Z$ operator ($\lambda^m = -1$), and denseness is lost. Using the $q$-deformed representation theory of SU(2) [36], with the elementary $R$-symbols parametrized by $q = e^{i2\pi/(k+2)}$, we obtain:

$$\lambda^m = (-1)^m q^{-m} = e^{imk\pi/(k+2)}, \tag{5}$$

$\lambda^m$ is fixed by $k$ and $m$. Eq. (5) then implies that the $m$-fold multiple braiding of the SU(2)$_k$ anyon model fails to be dense on the Bloch sphere whenever $2mk/(k+2) \in \mathbb{Z}$, i.e., $mk/(k+2)$ is an integer or a half-integer. In Table I we summarize, for several values of $k$, the predicted values of $m$ within $2 \le m \le 9$ at which denseness is lost. It is worth emphasizing that for the SU(2)$_2$ and SU(2)$_4$ models, the braid group representations are finite in their entirety, so that multiple braiding fails to be dense for every $m$; our criterion, however, predicts failure only at $m$ = 2, 4, 6, 8 for SU(2)$_2$ and at $m$ = 3, 6, 9 for SU(2)$_4$. Table I therefore characterizes the degeneracy of $m$-fold multiple braiding within models that are otherwise generically dense, whereas the failure at $k \in \{2, 4\}$ is global, affecting all $m$.

**Table I.** Selected SU(2)$_k$ anyon models with $2 \le k \le 9$, the values of $m$ ($2 \le k \le 9$) at which the $m$-fold braiding loses denseness, and the corresponding values of $\lambda^m$.

| $k$ | $m$ | $\lambda^m$ |
|---|---|---|
| 2 | 2, 4, 6, 8 | $e^{i\pi}$, $e^{i2\pi}$, $e^{i3\pi}$, $e^{i4\pi}$ |
| 3 | 5 | $e^{i3\pi}$ |
| 4 | 3, 6, 9 | $e^{i2\pi}$, $e^{i4\pi}$, $e^{i6\pi}$, |
| 5 | 7 | $e^{i5\pi}$ |
| 6 | 4, 8 | $e^{i3\pi}$, $e^{i6\pi}$ |
| 7 | 9 | $e^{i7\pi}$ |
| 8 | 5 | $e^{i4\pi}$ |
| 9 | None | None |

For the two-qubit encoding in Fig. 1(b), where the basis states are $\{|NC\rangle, |\mathbf{1},\mathbf{1}\rangle, |\mathbf{1}, X_2\rangle, |X_2, \mathbf{1}\rangle, |X_2, X_2\rangle\}$, the corresponding two-qubit MEBM is given by:

$$\left(\sigma_1^{(6)}\right)^m = \left(R_2^{11}\right)^m \oplus \left(\left(\sigma_1^{(3)}\right)^m \otimes I_2\right), \tag{6}$$

$$\left(\sigma_2^{(6)}\right)^m = \left(R_2^{11}\right)^m \oplus \left(\left(\sigma_2^{(3)}\right)^m \otimes I_2\right), \tag{7}$$

$$\left(\sigma_4^{(6)}\right)^m = \left(R_2^{11}\right)^m \oplus \left(I_2 \otimes \left(\sigma_2^{(3)}\right)^m\right), \tag{8}$$

$$\left(\sigma_5^{(6)}\right)^m = \left(R_2^{11}\right)^m \oplus \left(I_2 \otimes \left(\sigma_1^{(3)}\right)^m\right), \tag{9}$$

where $\oplus$ denotes the direct sum, $\otimes$ represents the direct product, and $I_2$ is the $2\times 2$ identity matrix. The notation $\left(\sigma_i^{(6)}\right)^m$ (with $1 \le i \le 5$) represents the $m$ braiding of the $i$-th and ($i$+1)-th anyons in the 6-anyon two-qubit encoding.

The exact form of the two-qubit operator $\left(\sigma_3^{(6)}\right)^m$ is provided in Appendix A, and the detailed derivation of all MEBMs procedure are presented in Supplementary Material I.

In addition to direct derivation from the $F$-matrices and $R$-symbols, MEBMs can also be obtained by applying the corresponding EBM $m$ times in succession. Specifically, for the $SU(2)_3$ (Fibonacci) model, the MEBMs are constructed by $m$-fold application of the single-qubit EBMs from Eqs. (7) and (8) of Ref. [16] and the two-qubit EBMs given in Eqs. (8)–(12) of Ref. [35].

Realizing universal quantum computation via multiple-braiding of $SU(2)_k$ anyon models (with $k > 2$, $k \neq 4$) amounts to using the single- and two-qubit MEBMs based on the anyon configurations shown in Fig. 1 to synthesizing a universal gate set {$H$, $T$, CNOT} [37].

Standard single-qubit gates such as the Hadamard ($H$) and phase ($T$) gates cannot be implemented directly via only a few braiding operations in $SU(2)_k$ anyon models (with $k > 2$, $k \neq 4$). Denseness of the braiding operations on the Bloch sphere manifests itself in the fact that the Solovay–Kitaev Algorithm (SKA), built recursively on the MEBMs of Eqs. (2) and (3), compiles the single-qubit gate set {$H$, $T$} to arbitrary precision; loss of denseness, in turn, implies that the MEBMs fail to compile {$H$, $T$}.

We compile the single-qubit gates with the GA-enhanced Solovay–Kitaev algorithm of Ref. [20], in which a genetic algorithm replaces the exhaustive search at each recursion level of the SKA, with the 0-level candidate braidwords supplied directly by the GA. The advantage of this approach for compiling high-fidelity single-qubit gates has been demonstrated numerically. The details of this computational method were provided in our previous work [8,20,24,38].

### III. Numerical Results and Gate Compilation Performance

We employ the GA-enhanced SKA to compile high-fidelity *H* and *T* gates from the one-qubit MEBMs of $SU(2)_3$ anyon models. For two-qubit entangling gates, we use the GA search to approximate braidword belonging to the local equivalence class [CNOT] using the MEBMs of two-qubit. These numerical constructions demonstrate the synthesis of a universal gate set. The analysis of the parameters used in the calculations is presented in Supplementary Material II.

In Appendix C, using braidwords obtained from the MEBMs of the $SU(2)_3$ anyon model as examples, we demonstrate how even-order braiding enables a topologically equivalent reduction in the number of anyons that need to be actively manipulated.

#### A. Single-Qubit Gate Construction

To quantify the distance between a braidword (constructed from the MEBMs of $SU(2)_k$ anyon models) and a target single-qubit gate, we employ the global phase–invariant distance *d* [18]. This metric *d* is insensitive to the overall phase difference, which is physically unobservable in quantum computation, and its definition as follow [8,18,20,38,39]:

$$d\left(U_0,U\right)=\sqrt{1-\frac{\left|Tr\left(U_0U^{\dagger}\right)\right|}{2}}, \tag{10}$$

where $U_0$ denotes the braidword consist of MEBMs, *U* represents the standard single-qubit gate, the dagger symbol † indicates the conjugate transpose, and *Tr* denotes the trace.

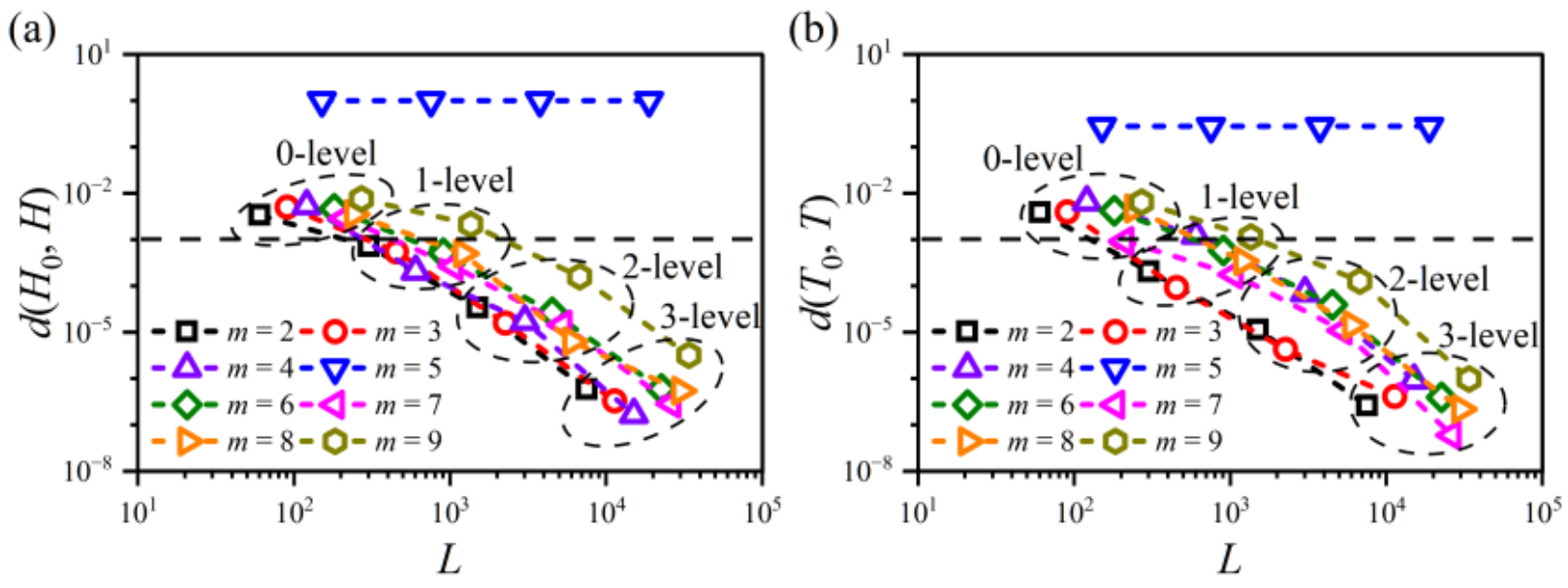

FIG. 2. Construction of standard (a) *H*-gate and (b) *T*-gate via multiple-braiding of $SU(2)_3$ anyon models using the GA-enhanced SKA. Each circle collects data points from different *m* at an identical approximation level *d*. The dashed line indicates the threshold $d = 10^{-3}$. Parameters are ($N = 1200$, $n = 800$, $P = 0.1$, $T = 600$, $G = 4000$).

As shown in Fig. 2, we employ the GA-enhanced SKA to compile standard *H* and *T* gates using both MEBMs ($m \in [2, 9]$) of the $SU(2)_3$ anyon models. The base length $l_0$ for the SKA 0-level approximation is set to $30^m$. In the multiple-braiding scheme, *m* successive braids are treated as one elementary operation unit; i.e., each MEBM corresponds to a $\left(\sigma_i^j\right)^m$ ($i$ = 1, 2 for $j$ = 3; $i$ = 1, 2, 3, 4, 5 for $j$ = 6). Consequently, for a multiple-braiding word composed of 30 MEBMs, the total number of *L* equals 30 × *m*. The 0-level braidwords for approximate the *H* and *T* gates obtained via GA search under the multiple-braiding of $SU(2)_3$ anyon scheme are summarized in Table II. Each 0- level braidword is obtained by running the GA three times and retaining the result with the smallest $d(U_0, U)$; the same three-run strategy is applied at every 0-level of approximation in the recursive SKA. We acknowledge that additional GA runs would yield smaller values of $d(U_0, U)$. However, the braidword length and the time cost at the $n$-level approximation of the SKA grow as $l_n = 5^n l_0$ and $t_n = 3^n t_0$, respectively [17]. Reaching the 3-level of approximation requires 27 0-level braidwords, each involving 81 GA runs in total, a cost further compounded by the need to combine braidwords of $m$ to compile both the $H$ and the $T$ gate. Three runs are therefore adopted as a compromise against this prohibitive time cost. For our purpose (numerically verifying the theoretical predictions of Table I), it suffices to demonstrate that for those $m$ predicted to be dense, the recursive GA-enhanced SKA substantially reduces the $d(U_0, U)$ to the ideal single-qubit gates (denseness), whereas for the remaining $m$ the ideal gates cannot be approximated (loss of denseness). The discrimination afforded by three runs is adequate for this purpose.

**Table** II. The 0-level braidwords and $d(U_0, U)$ metrics for approximate the $H/T$ gates based on multiple-braiding of $SU(2)_3$ anyon. $\{A, B, C, D\}$ corresponding to $\left\{\left(\sigma_1^{(3)}\right)^m, \left(\sigma_2^{(3)}\right)^m, \left(\left(\sigma_1^{(3)}\right)^m\right)^{-1}, \left(\left(\sigma_2^{(3)}\right)^m\right)^{-1}\right\}$.

| $m$ | Target gate | Braidwords | $d(U_0, U)$ |
|---|---|---|---|
| 2 | $H$ gate | CBABABABCBCBADABBCBABBCDCBADCD | 0.003369 |
| | $T$ gate | DDDDDCDABADABBBCDDCBCBCDCDCBBC | 0.003904 |
| 3 | $H$ gate | CCBCCDCBBABBBBABCCBBCCBCBCDAAD | 0.005059 |
| | $T$ gate | DDDAADDCCBBBCBADCDDCBBBABCCDAD | 0.003990 |
| 4 | $H$ gate | CBBABABBAABBAAAADAAADDADADABCB | 0.005415 |
| | $T$ gate | CBADADAADDADCCDDADDADDDDDDCDCC | 0.006545 |
| 5 | $H$ gate | BBBBBBBBBBBBBBBBBBBBBBBBBBBBAD | 1.000000 |
| | $T$ gate | DDDDDDDDDDDDDDDDDDDDDDDDDDDDDD | 0.275899 |
| 6 | $H$ gate | BBBBAADAADDDADADADCCCBCCBCBCBB | 0.004630 |
| | $T$ gate | ADADADAAABABBCCBABBCBBAAADADDA | 0.004227 |
| 7 | $H$ gate | CCDDDABCDCCCBBADDDDDCBAADABBBB | 0.002899 |
| | $T$ gate | CDAABBCDADCBCCBADAADDABBCCBCBA | 0.000931 |
| 8 | $H$ gate | BADCDABAADADCDAAAABABBCBCDADAB | 0.003533 |
| | $T$ gate | DABABCBBBBCCCDCCBCBCBABCBCBAAD | 0.004698 |
| 9 | $H$ gate | AAADAADCCCCDDDDDDADADDCCCDADCC | 0.007619 |
| | $T$ gate | CCDADADABCBBCCBCBCDADDADADCCBC | 0.006479 |

The computational results show that for $m \in [2, 9]$ (with $m$ a positive integer), only $m = 5$ fails to simultaneously construct high-precision $H$ and $T$ gates for $SU(2)_3$ anyon, while for all other values of $m$, increasing the approximation level of the GA-enhanced SKA leads to a systematic decrease in the $d(U_0, U)$, For $SU(2)_3$, $m = 5$ is the only multiple in $2 \le m \le 9$ that makes the single-qubit braiding representation "Pauli $Z$ operator" ( $\lambda^5 = e^{3\pi i} = -1,$ ); therefore, single-qubit $H$ and $T$ gates cannot be compiled.

This numerical result is consistent with the prediction of Table I. The threshold theorem for surface-code quantum error correction guarantees fault-tolerant operation as long as the physical error rate remains below approximately 1% [40,41]. The compilation error $d(U_0, U)$, however, is systematic in nature (independent of physical noise yet additive to it) and must therefore be kept well below this threshold, lest it dominate the total error budget. We accordingly impose the more stringent requirement $d(U_0, U) < 10^{-3}$, indicated by the black dashed lines in Fig. 2; this criterion, in line with common practice in topological quantum compilation [19], leaves ample headroom for physical noise, state-preparation and measurement errors. This requirement is already satisfied at the 2-level approximation for both gates, with typical $d(U_0, U)$ in the range $10^{-6} < d(H_0, H), d(T_0, T) < 10^{-3}$. If even higher accuracy is desired, the 3-level approximation can be reached, then the typical $d(U_0, U)$ reduce to the range $10^{-8} < d(H_0, H), d(T_0, T) < 10^{-5}$.

**B. Two-Qubit Entangling Gate and Leakage Error Analysis**

To establish that multiple-braiding in $SU(2)_k$ ($k > 2, k \neq 4$) anyon models is capable of universal quantum computation, besides the requirement that the set of single-qubit MEBMs generates arbitrary SU(2) operations, an entangling two-qubit gate must also be constructed. In this work, two-qubit MEBMs are used to approximate a locally equivalent [CNOT].

The computational approach to the approximate local equivalence class [CNOT] via braidword, together with the definition of the distance $d^{\mathrm{CNOT}}(A)$ between the braidword and the local equivalence class [CNOT], the $|M_{11}|$ (using to quantify leakage errors, arising from the eight off-block-diagonal elements of $B$ that couple the computational and non-computational subspaces, the leakage error of the braidword is negligible when $|1- |M_{11}|| < \delta$ for $\delta << 1$), and the unitary measurement $d^U$ are described detailedly in the references [8,26,38].

For convenience, we list below the defining equations used in the numerical calculations. The definition of the distance:

$$d^{\mathrm{CNOT}}(A) = \sum_{i=1}^{3} \Delta g_i^2, \Delta g_i = \left| g_i(A) - g_i(\mathrm{CNOT}) \right|, \tag{11}$$

where $A$ is a 4-dimensional matrix in the computational basis of two-qubit braidword, $g_1(\mathrm{CNOT}) = 0$, $g_2(\mathrm{CNOT}) = 0$, $g_3(\mathrm{CNOT}) = 1$. The definition of the unitary measurement:

$$d^U = Tr\left| A^\dagger A - \mathrm{I} \right|, \tag{12}$$

where I is a four-dimensional identity matrix.

Throughout the numerics, we impose the acceptance criteria $|M_{11}| > 0.95$ and $d^U < 0.1$, in line with those employed in Ref.[26] for compiling the local equivalence class [CNOT] with two-qubit braiding matrices of Fibonacci anyons.

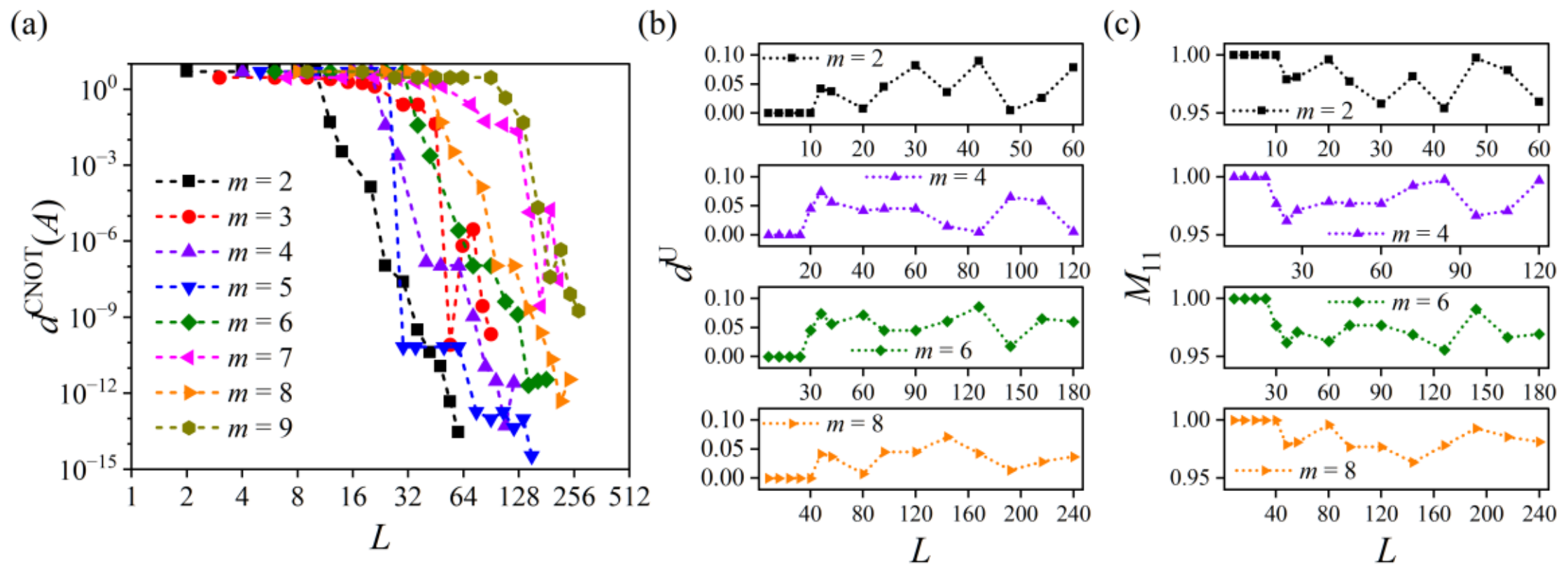

FIG. 3. Approximation of the local equivalence class [CNOT] via multiple-braiding of $SU(2)_3$ anyon models using the BF search ($L \le 7 \times m$) and GA ($L > 7 \times m$). Searches are constrained by the unitary measurement of $A$ $d^{\mathrm{U}} < 0.1$, and the $|M_{11}| > 0.95$. (a) Variation of $d^{\mathrm{CNOT}}(A)$ with $L$, (b) Variation of $d^{\mathrm{U}}$ with $L$ for even-order braiding, (c) Variation of $M_{11}$ with $L$ for even-order braiding. Parameters are ($N = 600$, $n = 300$, $P = 0.1$, $T = 300$, $G = 300$)

The results of approximating the local equivalence class [CNOT] using braidwords constructed from multiple-braiding operations in $SU(2)_3$ anyon models as shown in Fig. 3. The braidwords are obtained via BF search or GA.

The computational results demonstrate that increasing the braid length $L$ via GA systematically reduces the $d^{\mathrm{CNOT}}(A)$ to the target local equivalence class. Notably, for $L = 30 \times m$, all multiple-braiding ($2 \le m \le 9$) exhibits a significant advantage in approximating the local equivalence class [CNOT], achieving $d^{\mathrm{CNOT}}(A) < 10^{-7}$. For $L$ increased to $30 \times m$, detailed results (including the explicit braidwords, $d^{\mathrm{CNOT}}(A)$, $d^{\mathrm{U}}$, and $M_{11}$) obtained using MEBMs to approximate the local equivalence class [CNOT] for all $SU(2)_3$ models are summarized in Table III.

**Table III.** Braidwords of $SU(2)_3$ yielding the optimal $d^{\mathrm{CNOT}}(A)$, along with the corresponding $d^{\mathrm{U}}$ and $M_{11}$ were obtained via GA. $\{A, B, C, D, E, F, G, H, I, J\}$ corresponding to

$$\left\{\left(\sigma_1^{(6)}\right)^m, \left(\sigma_2^{(6)}\right)^m, \left(\sigma_3^{(6)}\right)^m, \left(\sigma_4^{(6)}\right)^m, \left(\sigma_5^{(6)}\right)^m, \left(\left(\sigma_1^{(6)}\right)^m\right)^{-1}, \left(\left(\sigma_2^{(6)}\right)^m\right)^{-1}, \left(\left(\sigma_3^{(6)}\right)^m\right)^{-1}, \left(\left(\sigma_4^{(6)}\right)^m\right)^{-1}, \left(\left(\sigma_5^{(6)}\right)^m\right)^{-1}\right\}.$$

| $m$ | Braidwords | $d^{\mathrm{CNOT}}(A)$ | $d^{\mathrm{U}}$ | $M_{11}$ |
|---|---|---|---|---|
| 2 | GHFCCCIAJDHFJFDDHFDFEDEDCJCIBB | $2.97088\times10^{-14}$ | 0.07879 | 0.95980 |
| 3 | EDJHHIIBEGCIEEAIHEAIHGDGAJGEDA | $2.20806\times10^{-10}$ | 0.07760 | 0.96042 |
| 4 | JJCDCJJDCAHFFDBJGCADEGDJBDHBAI | $2.67096\times10^{-12}$ | 0.00565 | 0.99717 |
| 5 | DCIAJBIECCIJDHHEIBCIJCDHDAJHDB | $3.4642\times10^{-15}$ | 0.01535 | 0.99229 |
| 6 | HEAGIGFBDFEEAHIGGADCBBJJFCJJDJ | $3.5819\times10^{-12}$ | 0.06013 | 0.96947 |
| 7 | BJFBBFDBCDAJJCDDJJHDCJADDCGDGF | $3.09586\times10^{-8}$ | 0.03589 | 0.98189 |
| 8 | AEFHFDCJCDFHFIFJHIEHJHHADCIJJA | $3.61483\times10^{-12}$ | 0.03687 | 0.98139 |
| 9 | CJBIJJIBFCBFGGECGFJAIFFGGFDHJH | $1.85709\times10^{-9}$ | 0.02063 | 0.98963 |

Although the compilation error $d^{\mathrm{CNOT}}(A)$ vanishes as the braid length $L$ grows as shown in Fig. 3, the GA search takes $d^{\mathrm{CNOT}}(A)$ alone as its objective, leaving the leakage indicator $M_{11}$ uncontrolled. A sizable leakage error implies that the braid operations readily drive the encoded states out of the computational subspace; a braidword with vanishing $d^{\mathrm{CNOT}}(A)$ but with $M_{11}$ deviating appreciably from unity is therefore of no practical use for quantum computation. We accordingly redefine the distance function as:

$$\varepsilon^{\mathrm{CNOT}} = a d^{\mathrm{CNOT}}(A) + b\left(1 - M_{11}^2\right), \tag{13}$$

where $a$ and $b$ weight the relative contributions of the gate error and the leakage error to the quality of a braidword, The leakage term $1 - M_{11}^2$ represents the weight of the output state lying outside the computational subspace [42,43]. The linear combination in Eq. (13) follows the same structural form as that in the framework of Niu et al. [44], in which gate infidelity and accumulated leakage are jointly optimized as soft penalty terms. This strategy has become a de facto standard in experimental studies [45,46].for simplicity we set $a = b = 1$ and rerun the GA with $\varepsilon^{\mathrm{CNOT}}$ as the objective. As shown in Fig. 4, $\varepsilon^{\mathrm{CNOT}}$ decreases steadily with increasing $L$: when the braid length is extended to $L = 30 \times m$, the multiple-braiding with $2 \le m \le 9$ reach $\varepsilon^{\mathrm{CNOT}} < 10^{-2}$. The corresponding braidwords are summarized in Table IV.

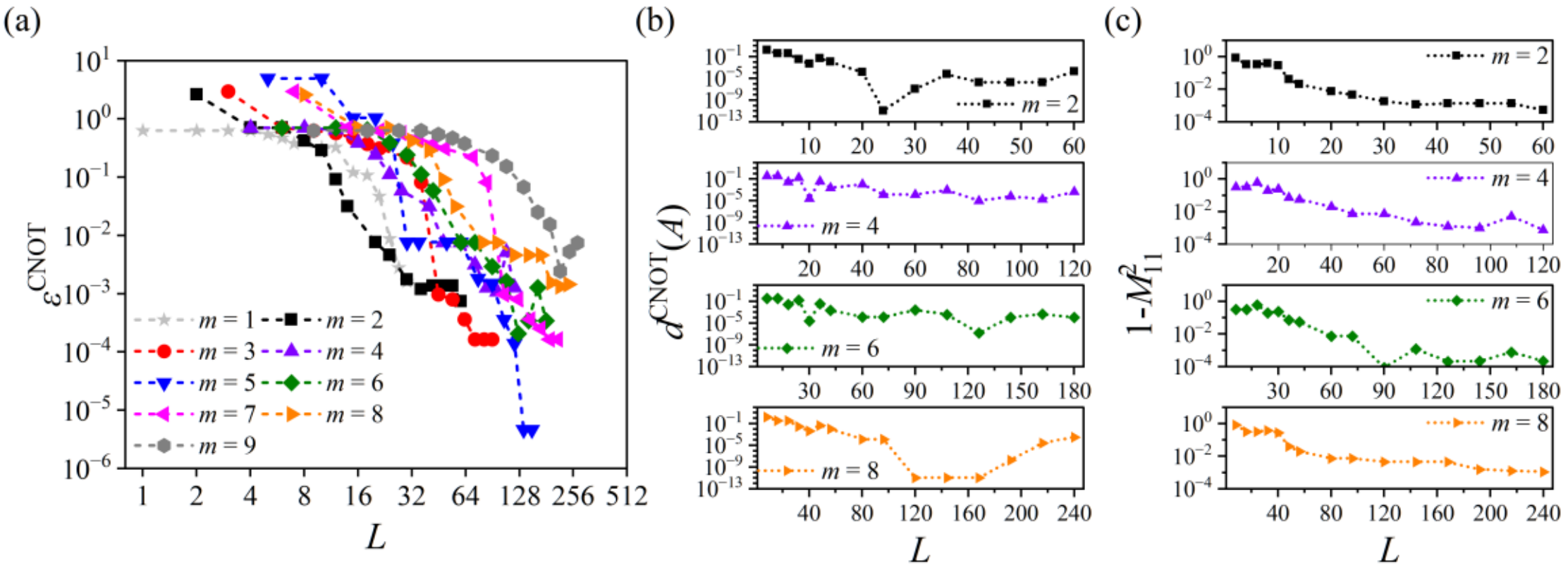

FIG. 4. Approximations of the $\varepsilon^{\mathrm{CNOT}}$ via multiple-braiding of SU(2)$_3$ anyon models using the BF search ($L \le 7 \times m$) and GA ($L > 7 \times m$). (a) Variation of $\varepsilon^{\mathrm{CNOT}}$ with $L$, (b) Variation of $d^{\mathrm{CNOT}}(A)$ with $L$ for even-order braiding, (c) Variation of $1 - M_{11}^2$ with $L$ for even-order braiding. Parameters are ($N = 4000$, $n = 2000$, $P = 0.1$, $T = 2000$, $G = 2000$).

As shown in Table III, the gate error is driven to below $10^{-7}$, while the leakage ($1 - M_{11}^2$) remains large: it ranges from $5.7 \times 10^{-3}$ ($m = 4$) to $7.9 \times 10^{-2}$ ($m = 2$), the latter corresponding to a leakage probability of nearly 8% per gate application. The large leakage observed in Table III stems from the objective function: the gate error $d^{\mathrm{CNOT}}(A)$ is explicitly minimized, whereas the leakage is merely bounded by the acceptance criterion ($|M_{11}| > 0.95$) and is never optimized itself. A gate error many orders of magnitude below every relevant physical scale represents wasted optimization budget: once $d^{\mathrm{CNOT}}(A) \le 10^{-4}$, further reduction leaves the total error, which is leakage dominated, essentially unchanged.

**Table IV.** Braidwords of SU(2)$_3$ yielding the optimal $\varepsilon^{\mathrm{CNOT}}$, along with the corresponding $d^{\mathrm{CNOT}}(A)$ and $1 - M_{11}^2$ were obtained via GA. $\{A, B, C, D, E, F, G, H, I, J\}$ corresponding

to $\left\{\left(\sigma_1^{(6)}\right)^m,\left(\sigma_2^{(6)}\right)^m,\left(\sigma_3^{(6)}\right)^m,\left(\sigma_4^{(6)}\right)^m,\left(\sigma_5^{(6)}\right)^m,\left(\left(\sigma_1^{(6)}\right)^m\right)^{-1},\left(\left(\sigma_2^{(6)}\right)^m\right)^{-1},\left(\left(\sigma_3^{(6)}\right)^m\right)^{-1},\left(\left(\sigma_4^{(6)}\right)^m\right)^{-1},\left(\left(\sigma_5^{(6)}\right)^m\right)^{-1}\right\}$.

| $m$ | Braidwords | $\varepsilon^{\mathrm{CNOT}}$ | $d^{\mathrm{CNOT}}(A)$ | $1 - M_{11}^2$ |
|---|---|---|---|---|
| 1 | ICEIEIEICIEIIFIFIIEICICEIEICCI | $1.61631\times10^{-3}$ | $1.33222\times10^{-9}$ | $1.61631\times10^{-3}$ |
| 2 | HHIEIHHIJFIIJIIIIJAJIHHIEIHHFG | $7.45300\times10^{-4}$ | $2.07338\times10^{-4}$ | $5.37962\times10^{-4}$ |
| 3 | EAAGAGHGAAGAGAAGFHGAAGAGAAGEHA | $1.66296\times10^{-4}$ | $1.17348\times10^{-4}$ | $4.89480\times10^{-5}$ |
| 4 | EHGCCAGHEABJABBAAAABBBHGCCAGHE | $1.28780\times10^{-3}$ | $5.05574\times10^{-4}$ | $7.82221\times10^{-4}$ |
| 5 | CBHFFBFCBBBFCBCABACBCABBBBBCBC | $4.67331\times10^{-6}$ | $9.31269\times10^{-24}$ | $4.67331\times10^{-6}$ |
| 6 | IFDDDDFHBBJCBHFGFJHBFGEGGHBBIF | $3.46896\times10^{-4}$ | $1.25186\times10^{-4}$ | $2.21709\times10^{-4}$ |
| 7 | AEHAEIEEIEIEEIHJIEEIEIEEIHEAEE | $1.66296\times10^{-4}$ | $1.17348\times10^{-4}$ | $4.89480\times10^{-5}$ |
| 8 | DDEBCGGFGGCEAAABABBAAEBCGGFGGC | $1.46273\times10^{-3}$ | $3.36120\times10^{-4}$ | $1.12661\times10^{-3}$ |
| 9 | DCCBBBCCDAABBEAEAABDBAACCBBBCC | $7.53602\times10^{-3}$ | $2.89645\times10^{-5}$ | $7.50705\times10^{-3}$ |

Since Tables III and IV are obtained at the same $L$ ($30 \times m$), they admit a direct comparison. The leakage is suppressed for every $m$, by factors ranging from 2.7 ($m$ = 9) to $3.3 \times 10^3$ ($m$ = 5), the price is a gate error that rises from machine precision to at most $5.1 \times 10^{-3}$ ($m$ = 4), and negligible compared with the leakage it buys down. The order $m$ = 5 deserves a separate remark, for it is at once the order compiled best and the order that cannot compile the single-qubit gates. It delivers the best compilation of the local-equivalence class [CNOT] under both objectives: the smallest of $d^{\mathrm{CNOT}}(A)$ (Fig. 3), while under the balanced objective the compilation is exact to machine precision and both $\varepsilon^{\mathrm{CNOT}}$ and the leakage attain their smallest values (Fig. 4). On the other hand, $m$ = 5 is precisely the order at which the recursive GA-enhanced SKA stagnates and fails to approximate $H$ and $T$ (Fig. 2). Compilation for [CNOT] only requires matching three real local invariants and does not require a dense representation. These numerical findings confirm that multiple-braiding ($2 \le m \le 9$, $m \ne 5$) in $SU(2)_3$ anyon models provides an effective means of realizing a high-fidelity universal quantum gates.

In Appendix B, we extend the numerical verification to the $SU(2)_5$, $SU(2)_6$, and $SU(2)_7$ anyon models, where the universal gate set {$H$, $T$, CNOT} is compiled from the $m$-fold MEBMs, testing the prediction of the theoretical criterion of Sec. II.

### C. Benchmarking gate compilation performance for Multiple-Braiding

To assess the compilation performance of multiple-braiding, we compare the $m$ = 1-6 on the universal gate set {$H$, $T$, CNOT}. A common word length must be a multiple of 1, … ,6 whose least common multiple is 60; the 0-level length is therefore set to $L$ = 120 (twice of 60), so that the $m$ = 6 retains 20 elementary blocks (MEBMs), while extending the comparison to $m$ = 7 would raise the common multiple to 420 at prohibitive computational cost. The $H$ and $T$ gates are compiled with the GA-enhanced SKA, each entry being the mean over 50 independent runs, with recursion 0, 1, 2, 3-levels corresponding to $L$ = 120, 600, 3000, and 15000; the CNOT class is compiled by a direct GA (20 runs) at $L$ = 60, 120, 180, 240.

All orders except $m$ = 5 behave alike as shown in Fig. 5: the $d(H_0, H)$ and $d(T_0, T)$ decrease steeply and monotonically with the recursion level, from $\backsim 10^{-2}$ at 0-level to $5 \times 10^{-7} \backsim 10^{-5}$ at 3-level (An overall gain of 3 to 4 orders of magnitude). The order $m$ = 3 is the most efficient throughout ($d(H_0, H) = 5.3 \times 10^{-7}$ and $d(T_0, T) = 5.8 \times 10^{-7}$) while consuming only one third as many elementary blocks as $m$ = 1 at equal $L$: the block structure of the intermediate orders purchases algebraic expressivity per block. In sharp contrast, at $m$ = 5 the SKA stagnates, at every level, at run-independent floor values: $d(H_0, H)$ =1 (the maximum allowed by the distance function, i.e., the best words found have vanishing trace overlap with $H$), and $d(T_0, T) = 0.2759$. That 50 independent runs at each of four levels return the same constant is the signature of a discrete braid-group image: the effective search space is finite, the GA reliably exhausts its best elements, and no amount of recursion improves the approximation, the numerical fingerprint of the non-denseness predicted by Table I through $2mk/(k+2) = 6 \in \mathbb{Z}$.

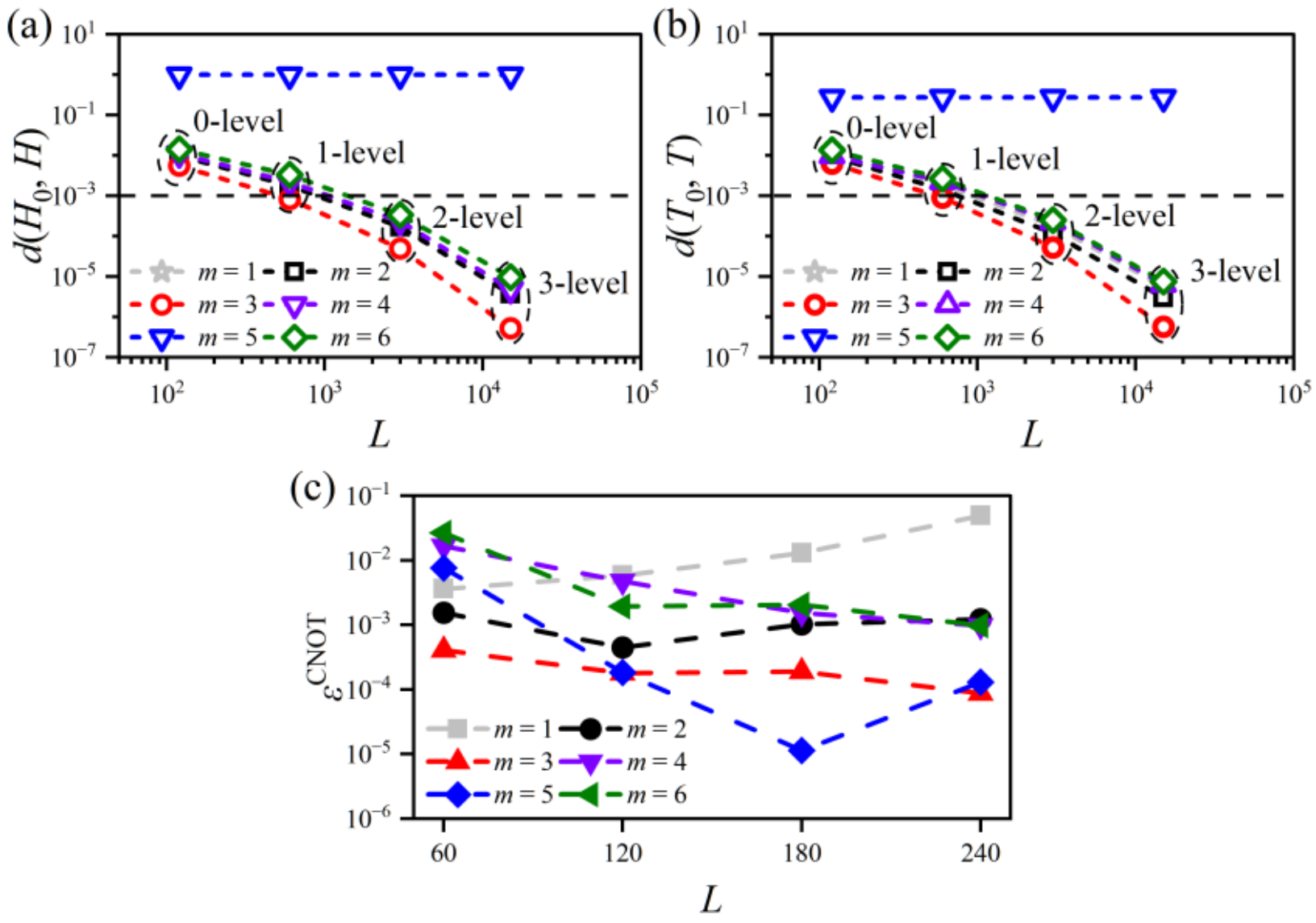

FIG. 5. Benchmarking gate compilation performance for multiple-braiding based on $SU(2)_3$ anyon ($1 \leq m \leq 6$). (a) Variation of $d(H_0, H)$ with $L$, (b) Variation of $d(T_0, T)$ with $L$, (c) Variation of $\varepsilon^{\mathrm{CNOT}}$ with $L$.

The direct-GA results for the CNOT class are fully consistent with this picture. At equal budget, $m$ = 5 again delivers the best compilation ($\varepsilon^{\mathrm{CNOT}} = 1.15 \times 10^{-5}$ at $L$ =180). The order $m$ = 3 is the runner-up ($\varepsilon^{\mathrm{CNOT}} = 8.8 \times 10^{-5}$ at $L$ =240). Short lengths favor the small orders ($m$ = 1-3 at $L$ =60), and the optimum shifts to larger orders as $L$ grows—the relevant resource is the number of elementary blocks $L/m$. It must be stressed that the error of $m$ = 1 grows monotonically with $L$ ($3.6 \times 10^{-3} \rightarrow 5.0 \times 10^{-2}$) and that of $m$ = 5 deteriorates beyond $L$ = 180: this is an optimizer artifact—a fixed GA budget constrains a search space expanding as $g^L$ ($g$ = 6, 10, $g$ is the type of MEBMs.)—rather than an intrinsic limitation of the multiple-braiding representation itself..

## IV Conclusion and Future Prospects

In this work we have generalized double-braiding of $SU(2)_k$ anyon models to multiple-braiding (the consecutive application of $m$ identical braid operations regarded as a single effective unit), and established when multiple-braiding supports universal quantum computation. We derived the general form of the single- and two-qubit MEBMs, and drawing on the $q$-deformed representation theory of SU(2), attributed the denseness of the $m$-fold braiding on the Bloch sphere to the behavior of the eigenvalue ratio: whenever $mk/(k+2)$ is an integer or half-integer, $\lambda^m$ degenerates to $\pm 1$, the MEBMs collapse to a scalar or a Pauli-Z operator (Up to a global phase), and denseness is lost. Numerical analysis of the $SU(2)_{3,5,6,7}$ models for $m \in [2, 9]$ confirms that the multiplicities at which the braid-group representation fails to be dense coincide exactly with this criterion.

Wherever denseness holds, the GA-enhanced SKA compiles the single-qubit gates $\{H, T\}$ from the MEBMs to high precision: the design target $d(U_0, U) < 10^{-3}$ is met already at the 2-level, while the 3-level approximation reduces $d(U_0, U)$ by roughly two additional orders of magnitude. For the two-qubit entangling gate, we redefine the objective as the sum of the gate and leakage errors, $\varepsilon^{\mathrm{CNOT}} = d^{\mathrm{CNOT}}(A) + (1 - M_{11}^2)$. At the braid length $L = 30m$ every $m$ whose single-qubit multiple braiding is dense yields a braidword with $\varepsilon^{\mathrm{CNOT}} < 10^{-2}$ in the $SU(2)_k$ anyon models with $k = 3, 5, 6, 7$. The only one exceptions occur in $SU(2)_7$: $m = 6$, although dense, does not reach the threshold at this length. Even in this case, however, an extremely small gate error $d^{\mathrm{CNOT}}(A)$ is still attainable; the shortfall of $\varepsilon^{\mathrm{CNOT}}$ is instead dominated by the leakage error.

Furthermore, even-order braiding is topologically equivalent to winding one anyon once around another, so that single-qubit gates require moving only one anyon and two-qubit gates only three, the remainder staying at rest—a substantial relief of the practical difficulty of actively manipulating non-Abelian anyons in TQC platforms based on topological superconductors or fractional quantum Hall systems. Taken together, our results support the conjecture that in $SU(2)_k$ anyon models ($k > 2$, $k \neq 4$) multiple-braiding loses universality only at the sparse set of multiplicities predicted by $2mk/(k+2) \in \mathbb{Z}$, all other $m$ sufficing for universal quantum computation. Extending the verification to larger $k$ is straightforward: determine the fusion rules, obtain the $F$-matrices and $R$-symbols from the $q$-deformed theory, fix a $m$ to define the MEBMs, and test whether they approximate the universal gate set $\{H, T, \mathrm{CNOT}\}$ to high precision. Combined with the topological reduction of actively manipulated anyons offered by even-order braiding, multiple-braiding thus provides a physically accessible and computationally efficient route to TQC.

**Acknowledgment** This work is supported by the National Natural Science Foundation of China (Grant No. 11204261), College of Physics and Optoelectronic Engineering training program (Grant No. 10QDZ18), a Key Project of the Education Department of Hunan Province (Grant No. 19A471), Natural Science Foundation of Hunan Province (Grant No. 2018JJ2381), Natural Science Foundation of Hunan Province (Grant No. 2026JJ90084).

**Data Availability Statement** The data that support the findings of this article are openly available [47].

**Declarations**

**Conflict of interest** No potential conflict of interest was reported by the authors. All authors of this manuscript have read and approved the final version submitted, and contents of this manuscript have not been copyrighted or published previously and are not under consideration for publication elsewhere.

**Appendix A: Derivation of the Exact MEBMs $\left(\sigma_3^{(6)}\right)^m$**

We present the specific form of the MEBMs $\left(\sigma_3^{(6)}\right)^m$ in SU(2)$_k$ anyon models within the anyon configuration illustrated in Fig. 1(b).

Performing braiding operations on the third and fourth anyons within the two-qubit basis ($|NC\rangle = [1\ \ 0\ \ 0\ \ 0\ \ 0]^{\mathrm{T}}, |00\rangle = [0\ \ 1\ \ 0\ \ 0\ \ 0]^{\mathrm{T}}$, $|01\rangle = [0\ \ 0\ \ 1\ \ 0\ \ 0]^{\mathrm{T}}$, $|10\rangle = [0\ \ 0\ \ 0\ \ 1\ \ 0]^{\mathrm{T}}$, $|11\rangle = [0\ \ 0\ \ 0\ \ 0\ \ 1]^{\mathrm{T}}$) by applying $F$- and $R$-moves (Note that trivial $F$-moves are omitted for simplicity), yields:

$$\left(\sigma_3^{(6)}\right)^m |NC\rangle = \left(\left(F_{2;03}^{112}\right)^{-1}\left(R_0^{11}\right)^m F_{2;30}^{112} + \left(F_{2;23}^{112}\right)^{-1}\left(R_2^{11}\right)^m F_{2;32}^{112}\right)|NC\rangle + \left(\left(F_{2;01}^{112}\right)^{-1}\left(R_0^{11}\right)^m F_{2;30}^{112} + \left(F_{2;21}^{112}\right)^{-1}\left(R_2^{11}\right)^m F_{2;32}^{112}\right)|11\rangle \quad (14)$$

$$\left(\sigma_3^{(6)}\right)^m |00\rangle = \left(R_0^{11}\right)^m |00\rangle \quad (15)$$

$$\left(\sigma_3^{(6)}\right)^m |01\rangle = \left(R_2^{11}\right)^m |01\rangle \quad (16)$$

$$\left(\sigma_3^{(6)}\right)^m |10\rangle = \left(R_2^{11}\right)^m |10\rangle \quad (17)$$

$$\left(\sigma_3^{(6)}\right)^m |11\rangle = \left(\left(F_{2;03}^{112}\right)^{-1}\left(R_0^{11}\right)^m F_{2;10}^{112} + \left(F_{2;23}^{112}\right)^{-1}\left(R_2^{11}\right)^m F_{2;12}^{112}\right)|NC\rangle + \left(\left(F_{2;01}^{112}\right)^{-1}\left(R_0^{11}\right)^m F_{2;10}^{112} + \left(F_{2;21}^{112}\right)^{-1}\left(R_2^{11}\right)^m F_{2;12}^{112}\right)|11\rangle \quad (18)$$

Eqs. (9) ~ (13) can be combined to yield $\left(\sigma_3^{(6)}\right)^m$. The derivation of the MEBMs is provided in Supplemental Material I.

**Appendix B: Single- and two-qubit gate compilation in higher-*k* models**

In this Appendix, we numerically verify the density criterion by compiling the universal gate set {$H$, $T$, CNOT} with the MEBMs of the SU(2)$_k$ models for $k$ = 5, 6, and 7.

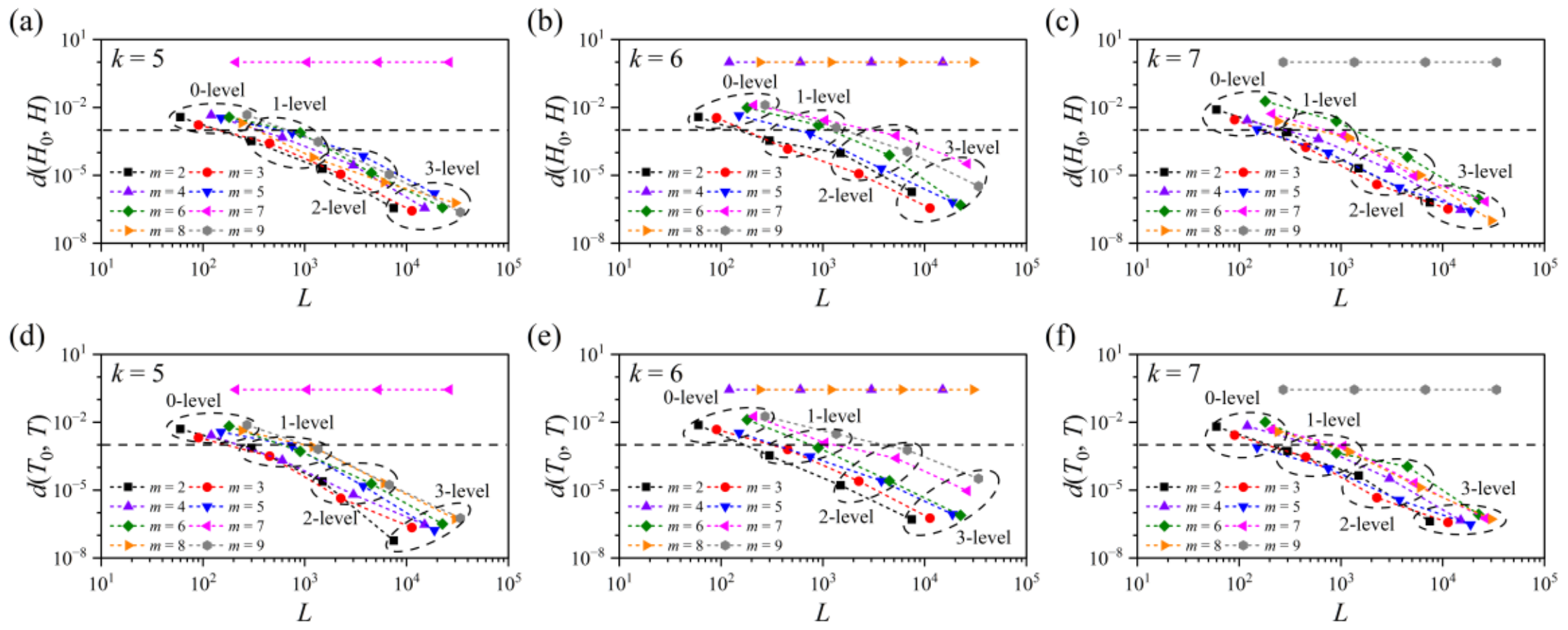

FIG. 6. Construction of standard (a)-(c) $H$-gate and (d)-(f) $T$-gate via multiple-braiding in the $SU(2)_5$, $SU(2)_6$, and $SU(2)_7$ anyon models using the GA-enhanced SKA. Each circle collects data points from different $m$ at an identical approximation level $d$. The dashed line indicates the threshold $d = 10^{-3}$.

Fig. 6 summarizes the GA-enhanced SKA compilation of the $H$ and $T$ gates in the $SU(2)_5$, $SU(2)_6$, and $SU(2)_7$ models, with $l_0 = 30m$ ($m = 2, \ldots, 9$). For all 20 multiplicities satisfying the density criterion $2mk/(k+2) \in \mathbb{Z}$ ($k = 5$: $m = 2$–6, 8, 9; $k = 6$: $m = 2$, 3, 5, 6, 7, 9; $k = 7$: $m = 2$–8), $d(U_0, U)$ decreases monotonically with each recursion level: the 0-level GA braidwords yield $d(U_0, U) \approx 10^{-3}$–$10^{-2}$, while after 3 recursion levels all 40 cases (both gates) fall below $3.3\times10^{-5}$, and most in the $10^{-7}$ range (best: $5.8\times10^{-8}$ for $T$ at $k = 5$, $m = 2$), an overall improvement of up to nearly five orders of magnitude, demonstrating that the GA-enhanced SKA compresses d dramatically in the higher-k models as well. In stark contrast, the four multiplicities predicted by the criterion to lose density ($m = 7$ of $k = 5$, $m = 4$ and $m = 8$ of $k = 6$, and $m = 9$ of $k = 7$) fail completely: $d(H_0, H)$ remains exactly 1 and $d(T_0, T)$ exactly 0.2759, unchanged at every recursion level. The $m$-fold braids generate pure phases on the computational space, so no braidword of any length can transcend this geometric bound, the failure reflects the structural collapse of the representation rather than a deficiency of the GA search, whose saturation of the bound already at 0-level, and its persistence thereafter, attests to the reliability of the search. All the computational results, together with the $k = 3$ observation in the main text, constitute the numerical verification of the density criterion.

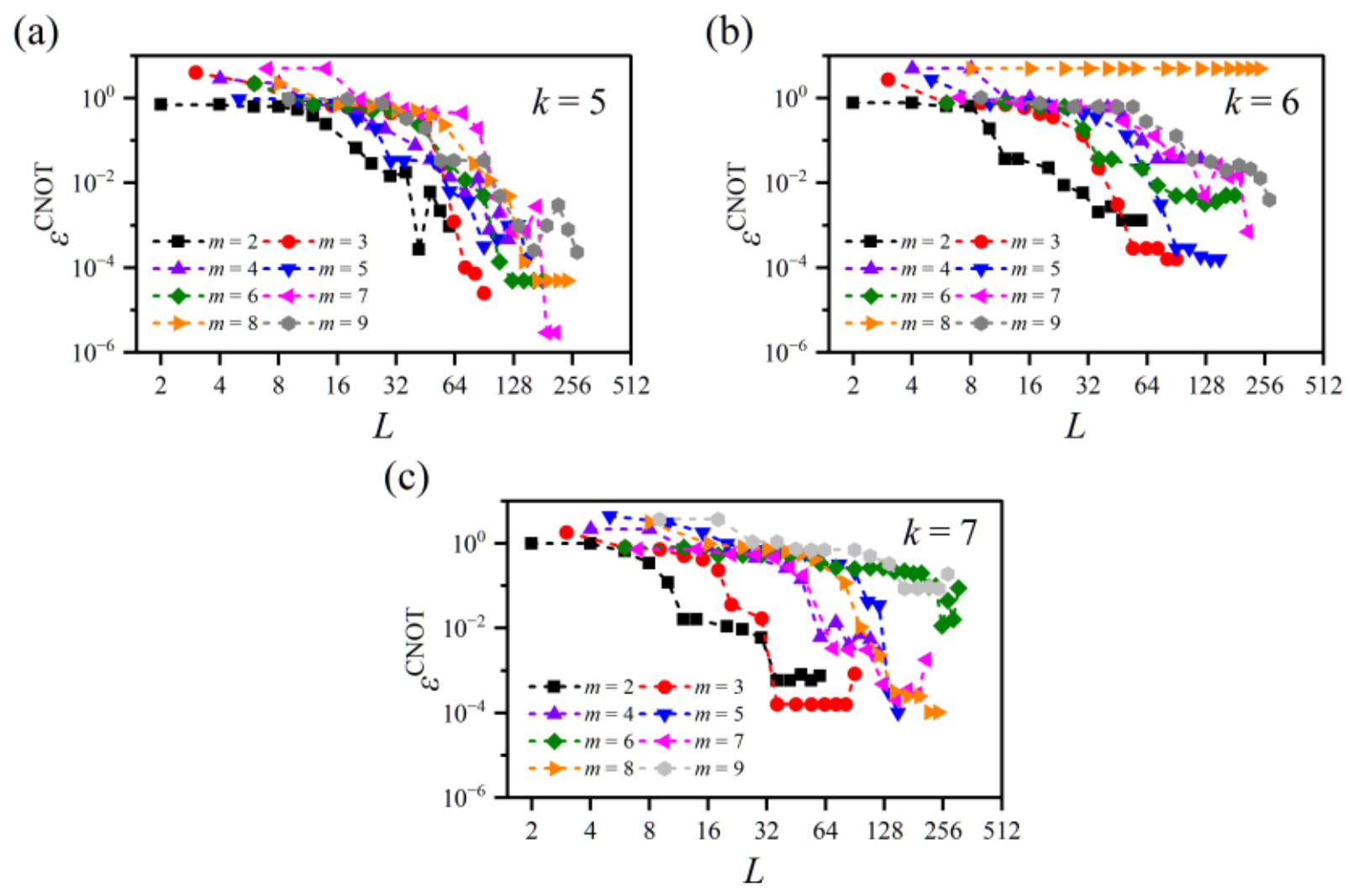

FIG. 7. Approximations of the $\varepsilon^{\mathrm{CNOT}}$ via multiple-braiding of (a) $SU(2)_5$ (b) $SU(2)_6$ (c) $SU(2)_7$ anyon models using the BF search ($L \leq 7 \times m$) and GA ($L > 7 \times m$).

Fig. 7 summarizes the GA compilation of the two-qubit CNOT gate (The distance is measured in $\varepsilon^{\mathrm{CNOT}}$) in the $SU(2)_5$, $SU(2)_6$, and $SU(2)_7$ models: for each $m = 2,\ldots,9$, the braidword length is scanned as $L = c \cdot m$ ($c$ = 1–7, 10–30), with $\varepsilon^{\mathrm{CNOT}}$ decreasing monotonically with $L$ and saturating at each ($k$, $m$). Of the 24 ($k$, $m$) combinations, 20 exhibit convergence, $\varepsilon^{\mathrm{CNOT}}$ starts at 0.7–5 for the shortest words ($L = m$) and drops to $10^{-4}$–$10^{-6}$ at the largest scanned lengths, an improvement of 3–6 orders of magnitude, with the best value of $2.98 \times 10^{-6}$ attained at $k = 5$, $m = 7$, demonstrating the significant compression of the two-qubit gate error by the GA. In contrast, four $m$s display anomalous behavior: at $k = 6$, $m = 8$, $\varepsilon^{\mathrm{CNOT}}$ remains at the saturation value 5 for all scanned lengths, corresponding to a complete failure of the compilation; at $k = 6$, $m = 4$, $\varepsilon^{\mathrm{CNOT}}$ stagnates at 0.037; at $k = 7$, $m = 6$, the convergence is slow ($\varepsilon^{\mathrm{CNOT}} = 0.19$ even at $L = 180$), even if the $L$ is increased to 306, $\varepsilon^{\mathrm{CNOT}}$ decreases only to $10^{-2}$, which indicates considerable leakage errors in the braiding operations for this case.; and at $k = 7$, $m = 9$, $\varepsilon^{\mathrm{CNOT}}$ decreases slowly, reaching $8.8 \times 10^{-2}$ only under an extended scan up to $L = 243$. Compared with the single-qubit criterion: the two non-dense multiplicities of $k = 6$ ($m$=4, 8) likewise degrade in the CNOT compilation—stagnation and complete failure, respectively—consistent in direction with the single-qubit case; by contrast, $k = 5$ and $m = 7$, non-dense at the single-qubit level, converges to $2.98 \times 10^{-6}$ in the CNOT compilation, while $k = 7$ and $m = 6$, dense at the single-qubit level, converges slowly. This indicates that the image of the two-qubit MEBMs in the four-dimensional computational space possesses a density structure independent of the single-qubit MEBMs. The convergence at $k = 5$, $m = 7$ is particularly noteworthy: even when the single-qubit $m$-fold braid image collapses to pure phases, the two-qubit representation can still support high-precision CNOT compilation, so the two-qubit computational resource is not constrained by the single-qubit density criterion.

## Appendix C: Reduction of Active Anyon Manipulation in Even-order Braiding

We now illustrate how even-order braiding topologically reduces the number of non-Abelian anyons that require active manipulation, using braidwords obtained from the SU(2)$_3$ model as concrete examples.

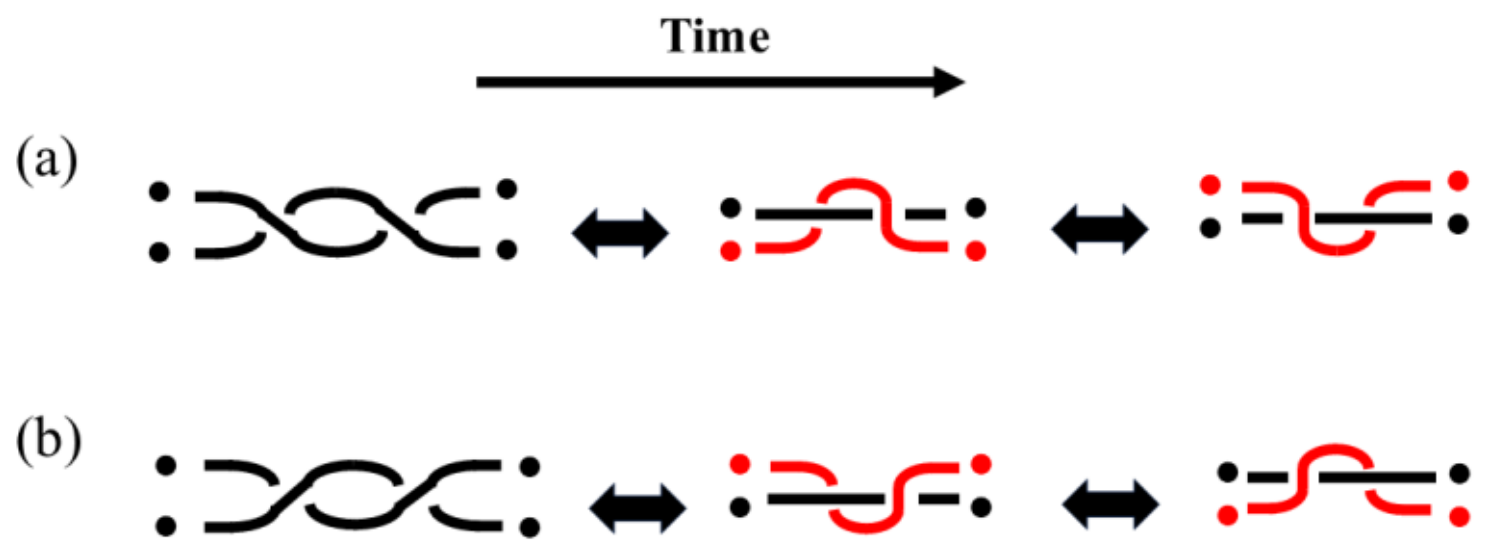

FIG. 8. The even-order braiding ($m$ = 2, equivalent to a double-braiding) of two anyons is topologically equivalent to one anyon making a full exchange around the other. Cases shown are for (a) clockwise and (b) counterclockwise double-braiding. Continuous double-braiding can naturally form even-order braiding.

Fig. 8(a) shows that performing two consecutive clockwise braids between two anyons is topologically equivalent to winding one anyon around the other by a full circle; the counterclockwise case is shown in Fig. 8(b).

Even-order braiding can reduce the number of non-Abelian anyons that require active physical manipulation through the topologically equivalent transformation illustrated in Fig. 8, where even-order braiding ($m$ = 2) is used as an example. Based on even-order braiding ($m$ = 2) in the SU(2)$_3$ anyon model, we employ the GA to search for braidwords that construct the standard universal quantum gate set {$H$, $T$, CNOT}. When the number of MEBMs is set to 15 (corresponding to $L$ =30), the results are summarized in Table V.

**Table V.** Braidwords and $d(U_0, U)$ for approximating $H$/$T$ gates (single-qubit) and the local equivalence class [CNOT] (two-qubit) using GA in even-order braiding ($m$ = 2) of SU(2)$_3$ anyon models. Single-qubit MEBMs: $\{A, B, C, D\}$ corresponding to $\left\{\left(\sigma_1^{(3)}\right)^2, \left(\sigma_2^{(3)}\right)^2, \left(\left(\sigma_1^{(3)}\right)^2\right)^{-1}, \left(\left(\sigma_2^{(3)}\right)^2\right)^{-1}\right\}$. Two-qubit MEBMs: $\{A, B, C, D, E, F, G, H, I, J\}$ corresponding to $\left\{\left(\sigma_1^{(6)}\right)^2, \left(\sigma_2^{(6)}\right)^2, \left(\sigma_3^{(6)}\right)^2, \left(\sigma_4^{(6)}\right)^2, \left(\sigma_5^{(6)}\right)^2, \left(\left(\sigma_1^{(6)}\right)^2\right)^{-1}, \left(\left(\sigma_2^{(6)}\right)^2\right)^{-1}, \left(\left(\sigma_3^{(6)}\right)^2\right)^{-1}, \left(\left(\sigma_4^{(6)}\right)^2\right)^{-1}, \left(\left(\sigma_5^{(6)}\right)^2\right)^{-1}\right\}$

| Target gate | $m$ | Braidwords | $d(U_0, U)$ | | |
|---|---|---|---|---|---|
| $H$ | 2 | CDABCDDADCBCBCC | 0.004643 | | |
| $T$ | 2 | DAABADADADCBADD | 0.008598 | | |
| | | | $\varepsilon^{\text{CNOT}}$ | $d^{\text{CNOT}}(A)$ | $1-M_{11}^2$ |
| CNOT | 2 | HIFIFJJCIJIIHHF | 0.004631 | $1.153774\times10^{-11}$ | 0.004631 |

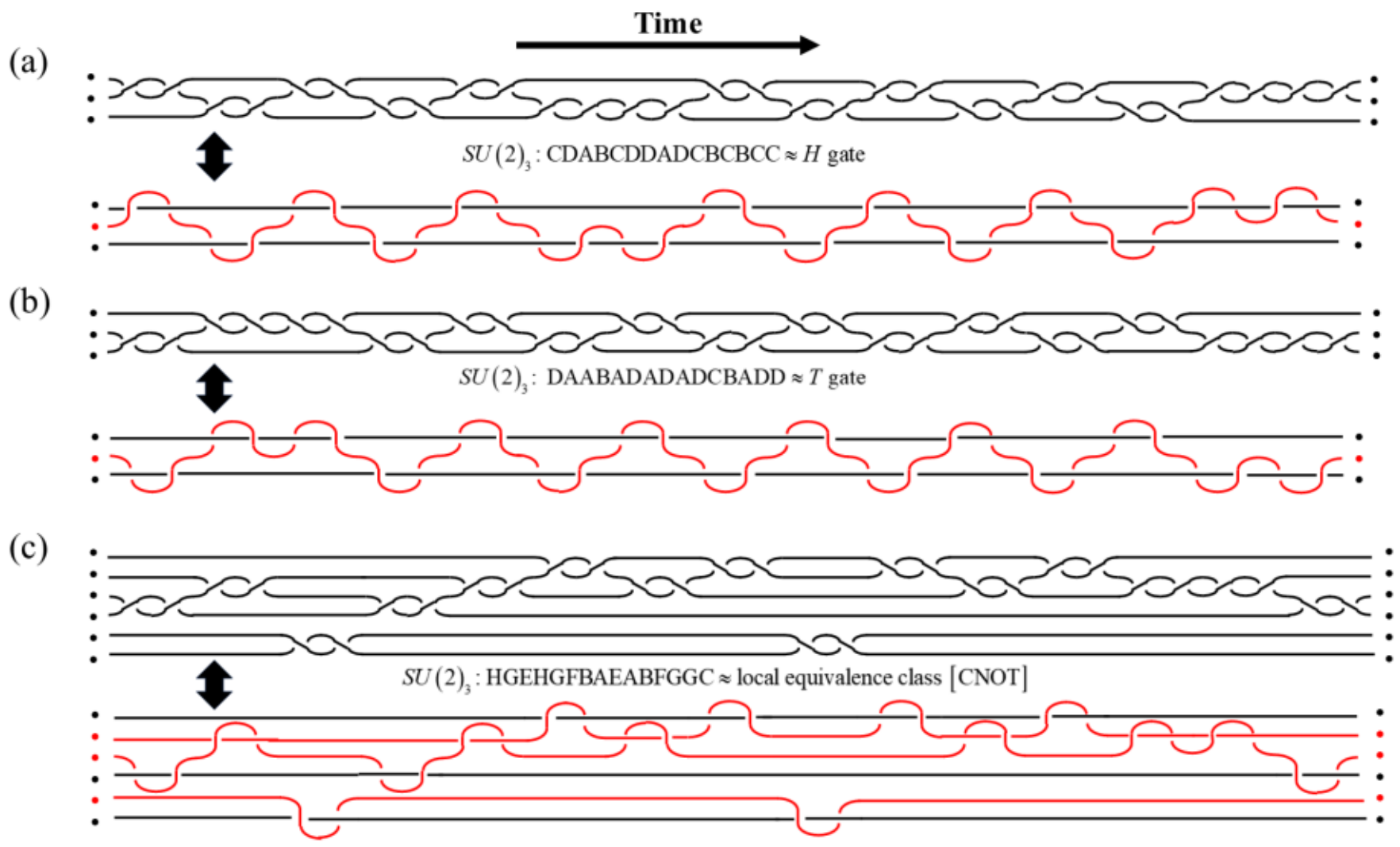

FIG. 9. Worldlines of braidwords approximating target gates. Braidwords for (a) the *H* gate, (b) the *T* gate (both up to a global phase), and (c) the local equivalence class [CNOT] were found via the GA using MEBMs ($m = 2$) of $SU(2)_3$ anyons. For each panel, the upper and lower worldline depictions are topologically equivalent. The anyon to be moved in the lower depiction is highlighted in red.

By applying the local topological transformations depicted in Fig. 8, the worldlines corresponding to the MEBM-derived braidwords for the $SU(2)_3$ model (explicitly, the braidwords of *H*, *T* and CNOT in Table V) can be equivalently transformed from the configurations shown in the upper panels of Fig. 9 to those in the lower panels, which involve moving fewer anyons. Specifically, for single-qubit braiding as shown in Fig .9 (a) and Fig .9 (b), only the second anyon (counting from top to bottom) needs to be moved around the first and third anyons, while the others remain stationary. For two-qubit braiding, it suffices to move only the second, third, and fourth anyons; all others can be kept at rest.